\documentclass[twocolumn]{autart}

\usepackage{graphicx}

	\usepackage{amssymb,amsfonts}
	\usepackage[cmex10]{amsmath}

	\usepackage{graphicx}
	\usepackage{amsfonts}
\usepackage{enumerate} 
	\usepackage[all,cmtip]{xy}
	\usepackage{tikz}
\usepackage[numbers,sort&compress]{natbib}
\usepackage{subfigure}
\usepackage{multirow} 
\usepackage{graphicx}
\usepackage{epsfig}
\usetikzlibrary{calc}
\usepackage{array,arydshln}
\usepackage{blkarray}
\usepackage{array}
\usepackage{cases}
\usepackage{mathrsfs}
\usepackage{enumerate}
\usepackage[colorlinks,linkcolor=red,anchorcolor=blue,citecolor=blue]{hyperref}

\newtheorem{theorem}{\textbf{Theorem}}

\newtheorem{lemma}{\textbf{Lemma}}
\newtheorem{proposition}{\textbf{Proposition}}
\newtheorem{remark}{\textbf{Remark}}
\newtheorem{example}{\textbf{Example}}

\newtheorem{definition}{\textbf{Definition}}

\journal{Journal Name}

\begin{document}

\begin{frontmatter}
\endNoHyper

\title{Distributed adaptive   stabilization \thanksref{footnoteinfo}} 

\thanks[footnoteinfo]{
		Corresponding author: Zhongkui Li}

\author[TUE]{Zhiyong Sun} \ead{z.sun@tue.nl, sun.zhiyong.cn@gmail.com},   
\author[LTH]{Anders Rantzer} \ead{anders.rantzer@control.lth.se},
\author[PKU]{Zhongkui Li} \ead{zhongkli@pku.edu.cn}, 
\author[LTH]{Anders Robertsson} \ead{anders.robertsson@control.lth.se},

\address[TUE]{Department of Electrical Engineering, Eindhoven University of Technology (TU/e), Eindhoven, the Netherlands.}
\address[LTH]{Department of Automatic Control, Lund University, Sweden.}
\address[PKU]{State Key Laboratory for Turbulence and Complex Systems, Department of Mechanics and Engineering Science, College of Engineering, Peking University, Beijing, China.}  



\begin{abstract}
In this paper we consider distributed adaptive stabilization for uncertain multivariable linear systems with a time-varying diagonal \textit{matrix} gain. We show that uncertain multivariable linear systems are stabilizable by diagonal matrix high gains   if the system matrix is an H-matrix with positive diagonal entries. Based on matrix measure and stability theory for diagonally dominant  systems, we consider two classes of uncertain linear systems, and derive a threshold condition to  ensure their exponential stability   by a monotonically increasing diagonal gain matrix.  When each individual gain function  in the matrix gain  is updated by state-dependent functions using only local state information,   the boundedness and convergence of both system states and adaptive matrix gains are guaranteed.  We apply the adaptive distributed   stabilization approach  to adaptive synchronization control for large-scale complex networks consisting of nonlinear node dynamics and  time-varying  coupling weights.  A unified framework  for adaptive synchronization is proposed that includes several general design approaches for adaptive coupling weights to guarantee network synchronization. 
\end{abstract}


\end{frontmatter}


\section{Background and motivations}

\subsection{Adaptive stabilization control}
Adaptive stabilization is a popular approach in direct adaptive control for stabilizing  uncertain plants with unknown or uncertain system parameters. Different to   estimation-based indirect adaptive control methods, adaptive stabilization  aims  to stabilize uncertain systems without estimating or identifying  system parameters during the stabilization process. A simple method for direct adaptive control for uncertain plants is   adaptive high-gain stabilization. To start with, 
consider the following stabilization problem for a time-invariant system in $\mathbb{R}^n$:
\begin{align}
    \dot x(t) =  Ax(t) + B u(t), y(t) = Cx(t),
\end{align}
where the system matrices $A, B$ and $C$ with proper dimensions are unknown, and the stabilization control law is designed as
\begin{align} \label{eq:scalar_gain}
    u(t) =  -k(t)y(t),
\end{align}
where $k(t)$ is a scalar \textit{time-varying} positive gain function. The adaptive stabilization approach is promising for stabilizing uncertain systems, as long as the system is \textit{high-gain stabilizable} \cite{Ilchmann}. We call an uncertain plant   high-gain stabilizable with the simple feedback of \eqref{eq:scalar_gain}, if the closed-loop system is stable under $k(t) > \bar k$ by a sufficiently large gain $\bar k$. Very often a lower bound estimation of the threshold gain value $\bar k$ depends on the system parameters and dynamics which might be unknown or uncertain. To this end, adaptive high-gain stabilization allows online adjustment of  adaptive gain tuning which eventually generates a feedback gain greater than the threshold gain $\bar k$ and therefore ensures system convergence. The first use of adaptive stabilization could be dated back to the 1970s (see, e.g.,  an early paper \cite{fradkov1976quadratic}). The literature on adaptive high-gain stabilization has seen a fruitful theoretic development as early as in the 1980s, with several key developments in some influential papers e.g., \cite{byrnes1984adaptive, maartensson1985order,nussbaum1983some, mudgett1985adaptive,ALLGOWER1997881}, and also successful   applications to industrial  control systems and uncertain mechatronic systems \cite{Ilchmann,hackl2012non}.

The seminal paper by  Byrnes and   Willems \cite{byrnes1984adaptive}  considered the following adaptive gain updating law
\begin{align} \label{eq:Byrnes_Willems_update}
    \dot k(t) = \|y(t)\|^2, k(0) = k_0 >0
\end{align}
and showed that, if the system is minimum phase and all eigenvalues of the system matrix product $-CB$  have negative real parts,  
then the time-varying linear system 
\begin{align} \label{eq:eq:Byrnes_Willems_matrix}
        \dot x(t) =  (A - k(t)BC) x(t)
\end{align}
is asymptotically stable. In addition, the scalar time-varying gain $k(t)$ updated by \eqref{eq:Byrnes_Willems_update} is upper bounded and convergent in the sense that $\text{lim}_{t\rightarrow \infty}k(t) = k_{\infty} <\infty$. The adaptive gain method has also been generalized to other types of dynamical control systems, such as tracking control systems \cite{ALLGOWER1997881} and nonlinear systems  (see e.g.,  \cite{lei2006universal}).  We refer the readers
to some recent surveys \cite{ilchmann2008high,barkana2016adaptive} for recent developments and applications of adaptive stabilization control.

\subsection{Distributed adaptive stabilization: problem formulation}
The main focus of this paper is to extend the   classical adaptive stabilization theory with time-varying \textit{scalar} gains to   adaptive \textit{matrix} gains and develop a \textit{distributed} adaptive stabilization theory. 
The adaptive stabilization approach in \eqref{eq:eq:Byrnes_Willems_matrix} under the scalar gain updating law \eqref{eq:Byrnes_Willems_update}, which involves all system states or outputs, is
not scalable for large-scale complex systems that often contain hundreds or thousands
of states. We therefore formulate the distributed adaptive stabilization
problem by replacing the  scalar gain function in \eqref{eq:eq:Byrnes_Willems_matrix}   by a matrix gain $K(t)$ in a diagonal  form $K(t) : = \text{diag}\{k_1(t), k_2(t), \cdots, k_n(t)\} \in \mathbb{R}^{n\times n}$, in the sense that each diagonal gain is associated with stabilization of each distributed channel of the uncertain system. Furthermore,  the updating of each diagonal gain element $k_i(t)$ should involve only  local information of the corresponding  state $x_i(t)$. 
Without loss of generality we assume in the sequel that $C = I_n$; i.e., the system output measures the states of the uncertain system. We will consider the following systems 
\begin{align}  \label{eq:system_example_I}
    \dot x(t) = (A - K(t)B)x(t),\,\,\,\,\,\,\text{System (I)}
\end{align}
and 
\begin{align}  \label{eq:system_example_II}
    \dot x(t) = (A - BK(t))x(t),\,\,\,\,\,\,\text{System (II)}
\end{align}
where $A$ and $B$ are unknown system matrices. Note that since the diagonal matrix $K(t)$ does not necessarily commute with the system matrix $B$, System (I) in \eqref{eq:system_example_I} and System (II) in \eqref{eq:system_example_II} need to be treated separately. \footnote{System (I) can be interpreted in the output feedback structure, while 
System (II) is related to the state feedback structure in linear system theory. For both systems, the diagonal  control gain $K(t)$ is associated with stabilization for each channel of the uncertain linear systems. A non-diagonal $K(t)$ does not meet the distributed stabilization requirement, which will be considered in future work.}  When the square matrix $B$ is non-singular, the system  \eqref{eq:system_example_I} is minimum phase \cite{Ilchmann}, and the Byrnes-Willems condition \cite{byrnes1984adaptive} for such uncertain systems to be scalar high-gain stabilizable reduces to that $-B$ is a Hurwitz matrix. However, for a time-varying matrix   gain $K(t)$ in  \eqref{eq:system_example_I} or  \eqref{eq:system_example_II}, we will show in this paper that some additional conditions on the matrix $B$ should be imposed to guarantee matrix high-gain stabilizability, which therefore demands new theories and tools  to stabilize uncertain systems of \eqref{eq:system_example_I} or  \eqref{eq:system_example_II}. We will also provide a general approach for adjusting adaptive matrix gains according to system state evolution in order to ensure system convergence and upper bounded gain matrix.

\subsection{Motivation and application: scalable and distributed control of networked systems}  
The motivation choice of considering a diagonal matrix gain function instead of a scalar gain function is inspired  by many recent applications in  \textit{distributed} control for multi-agent systems \cite{knorn2016overview}, and \textit{scalable}  control of large-scale networked systems  \cite{rantzer2018tutorial}.  In these control scenarios, very often each individual system is associated with some local gain functions while a global and uniform gain is inaccessible or very hard to compute. Each individual system should update its own adaptive gain function using local information to guarantee the stability of the overall system, while the local adaptive gain and its updating law are often different among all systems in a networked environment. Furthermore,  since individual adaptive gains are updated in a local and distributed way without involving any global information, the distributed adaptive stabilization approach is scalable and independent of the system size, which is a favorable property particularly applicable  for stabilization and regulation of large-scale networks.

A typical application   of distributed adaptive  stabilization with matrix high gains is adaptive synchronization control of complex networks. Synchronization   of complex network systems has a long history   \cite{wu1995synchronization}, \cite{nijmeijer1997observer}, \cite{dorfler2014synchronization}, which  has been motivated by vast applications that involve increasingly complicated inter-connected systems. The condition for reaching network synchronization depends on both individual system dynamics and the overall network topology, which are often hard to obtain and  also result in a high computational cost. Therefore, distributed adaptive   stabilization is a suitable approach for designing an adaptive synchronization protocol. In recent years there has been an increasing interest in the design of adaptive coupling gain tuning functions to ensure adaptive synchronization \cite{yu2013distributed, su2013decentralized, shafi2015adaptive}. In this paper, via the matrix high-gain approach, we will  provide a unified framework on adaptive synchronization and suggest several novel and general  approaches on designing local adaptive coupling functions to achieve   synchronization in complex network systems.

\subsection{Contributions  and paper organization}

The main contribution of this paper is a comprehensive study of distributed adaptive stabilization theory, while we also present several conditions on   matrix high-gain  stabilizability of  uncertain linear systems. Applications to nonlinear networked systems and adaptive synchronization will also be shown, which provide  novel insights on the general design of adaptive coupling weights to ensure network synchronization. 
A preliminary version was presented in \cite{sun2019adaptive}. In this paper we will generalize the matrix condition (while in \cite{sun2019adaptive} we focused on the M-matrix condition), and  will further present a complete study on distributed adaptive stabilization for both systems (I) and (II), based on some new applications of tools such as H-matrix and matrix measure theory. In the development of the main results, the paper will also prove some interesting results on exponential convergence of time-varying linear systems.  

This paper is organized as follows. Section~\ref{sec:preliminary} presents definitions and properties of some special matrices, and introduces several convergence results for diagonally dominant systems. Distributed adaptive  stabilization with matrix high gains for System (I) and System (II) is presented in Section~\ref{sec:main_matrix_gain} and Section~\ref{sec:main_matrix_gain_II}, respectively. We discuss applications of adaptive matrix gain stabilization to network synchronization in Section~\ref{sec:network_synchronization}, followed by conclusions in Section~\ref{sec:conclusions}. In appendix, we present some background on matrix measure and proofs.

\section{Definitions and preliminaries} \label{sec:preliminary}
\subsection{Notations}
The notations in this paper are fairly standard. For a real symmetric matrix $A$, we use $\text{min} \lambda(A)$   to denote its minimum eigenvalue, and  $A \prec 0$ to indicate that $A$ is  negative definite.  The notation $\rho(A)$ denotes the spectral radius of a square matrix $A$.   The null space of a real matrix $A$ is denoted by $\text{null}(A)$. The notation ${\bf{1}}_n$ denotes an $n$-vector with all $1$'s, and $I_n$ denotes an $n \times n$ identity matrix. The symbol $\otimes$ denotes Kronecker product.  For a vector $x \in \mathbb{R}^n$, the notation $\|x\|_1$ denotes the vector 1-norm, i.e., $\|x\|_1 = \sum_{i=1}^n |x_i|$, and  $\|x\|_\infty$ denotes the vector infinity norm, i.e., $\|x\|_\infty = \text{max}_{i=1,\cdots,n} |x_i|$. By default, the notation $\|x\|$ for a vector $x \in \mathbb{R}^n$ is interpreted as the 2-norm, unless otherwise specified.  

\subsection{M-matrix, H-matrix, and generalized  diagonally dominant matrix}
We present definitions of certain special matrices which will be frequently used in this paper. All matrices discussed in this paper are real-valued matrices. 
\begin{definition} \label{def:M_matrix} (\textbf{M-matrix})
     A matrix $A \in \mathbb{R}^{n \times n}$ is called an M-matrix, if its non-diagonal entries are non-positive and its  eigenvalues have positive real parts. \footnote{In this paper by convention an M-matrix is meant a \textit{non-singular} M-matrix.
Note that there is also a counterpart to non-singular M-matrix,
termed \textit{singular M-matrix} (A typical example of a singular M-matrix is the  graph Laplacian matrix~\cite{godsil2013algebraic}). This will be made clear in the context.}
\end{definition}
 
\begin{definition} \label{def:gene_row_DD}(\textbf{Generalized row-diagonally dominant matrix})
A matrix $A  = \{a_{ij}\}\in \mathbb{R}^{n \times n}$ is   generalized   \textit{\textbf{row}}-diagonally dominant, if there exists $x = (x_1, x_2, \cdots, x_n) \in \mathbb{R}^n$ with $x_i >0$, $\forall i$, such that
\begin{align}
    |a_{ii}| x_i > \sum_{j=1, j \neq i}^{n} |a_{ij}|x_j, \forall i = 1, 2, \cdots, n.
\end{align}
\end{definition}

\begin{definition} \label{def:gene_column_DD} (\textbf{Generalized column-diagonally dominant matrix})
A matrix $A  = \{a_{ij}\} \in \mathbb{R}^{n \times n}$ is   generalized   \textit{\textbf{column}}-diagonally dominant, if there exists $x = (x_1, x_2, \cdots, x_n) \in \mathbb{R}^n$ with $x_i >0$, $\forall i$, such that
\begin{align}
    |a_{jj}| x_j > \sum_{i=1, i \neq j}^{n} |a_{ij}|x_i, \forall j = 1, 2, \cdots, n.
\end{align}
\end{definition}

If the positive vector $x$ in the above definitions is chosen as the all-ones vector ${\bf{1}}_n$, then the two terms in Definitions~\ref{def:gene_row_DD} and \ref{def:gene_column_DD}  reduce to the conventional definitions of (strict) row-/column-diagonally dominant matrices.

\begin{definition} \label{def:comparison_matrix} (\textbf{Comparison matrix and $H$-matrix})
For a real matrix $A = \{a_{ij}\} \in \mathbb{R}^{n \times n}$, we associate it with a  comparison matrix $M_A = \{m_{ij}\} \in \mathbb{R}^{n \times n}$, defined by 
\begin{align}
    m_{ij} =   \left\{
       \begin{array}{cc}
       |a_{ij}|,  &\text{  if  } \,\,\,\,j  = i;  \\ \nonumber
       -|a_{ij}|,  &\text{  if  } \,\,\,\, j  \neq i.   \nonumber  
       \end{array}
      \right.
\end{align}
A given matrix $A$ is called an H-matrix if its comparison matrix $M_A$ is an M-matrix. 
\end{definition}

Apparently, the set of M-matrices is a subset of H-matrices. An iterative criterion for determining H-matrix can be found in \citep{bishan1998iterative}.  The generalized row-diagonally dominant matrix is also closely related to M-matrix. 
The following fact appears in \citep{fiedler1962matrices} (see also \citep[Chapter 2.5]{roger1994topics}). 
 
\begin{lemma} \label{lemma:H_matrix}
A given matrix $A \in \mathbb{R}^{n \times n}$ is an $H$-matrix if and only if there exists a positive diagonal matrix $D$ such that $AD$ is row-diagonally dominant. 
\end{lemma}
The Lemma actually states that  H-matrices and generalized row-diagonally  dominant matrices are the same.  
We shall prove the following more general result that shows the equivalence between  generalized row-diagonal dominance,   generalized column-diagonal dominance, and  H-matrices.  

\begin{theorem} \label{theorem:_H_matrix}
Given a matrix $A = \{a_{ij}\} \in \mathbb{R}^{n \times n}$, the following statements are equivalent. 
\begin{enumerate}[(i)]
 \label{claim:H_matrix}
    \item $A$ is an $H$-matrix;
    \item $A$ is generalized \textit{row}-diagonally dominant;
    \item There exists a positive diagonal matrix $\bar D = \text{diag}\{\bar d_1, \bar d_2, \cdots, \bar d_n\}$, such that $\bar D^{-1} A \bar D$ is row-diagonally dominant; 
    \item $A$ is generalized \textit{column}-diagonally dominant;
    \item There exists a positive diagonal matrix $\tilde D = \text{diag}\{\tilde d_1, \tilde d_2, \cdots, \tilde d_n\}$, such that $\tilde D A \tilde D^{-1}$ is column-diagonally dominant. 
\end{enumerate}
\end{theorem}

The proof is presented in Appendix. Applying the   Gershgorin circle theorem \cite{roger1994topics}, we immediately obtain the following result as a direct consequence of Theorem~\ref{theorem:_H_matrix}. 
\begin{proposition}
Let $B =\{b_{ij}\} \in \mathbb{R}^{n\times n}$ be an $H$-matrix.  Then $B$ is non-singular. Further suppose that $B$ has all positive diagonal entries, i.e., $b_{ii} >0, \forall i$. Then all of its eigenvalues have positive real parts; i.e.,   $-B$ is a non-singular Hurwitz matrix.  
\end{proposition}

\subsection{Exponential convergence of diagonally dominant time-varying  systems}
In this section, by applying the theory of matrix measures (see Appendix),   we develop some results on the solution bound and exponential convergence of time-varying linear systems with diagonally dominant system matrices. 

\begin{lemma} \label{lemma:row_dd_system}
(Row-diagonally dominant linear system)
Consider a time-varying linear system $\dot x(t) = A(t)x(t)$, where $A(t)$ is a continuous-time Hurwitz matrix with row-diagonally dominant entries $\forall t \geq t_0$. Then it holds that
\begin{align}
        \|x(t)\|_\infty \leq \|x(t_0)\|_\infty e^{\int_{t_0}^{t} \alpha_r(t') \text{d}t'},  \forall t \geq t_0, 
\end{align}
where $\alpha_r(t') = \text{max}_{i =1, 2, \cdots, n} \left(a_{ii}(t') +\sum_{j=1, j\neq i}^n |a_{ij}(t')|\right)$ and $\alpha_r(t')<0$.
\end{lemma}

\textbf{Proof}
Applying Lemma~\ref{theorem:measure_exponential}, and choosing the vector norm as the infinity norm with the matrix measure induced by infinity vector norm in \eqref{eq:measure_infinity_norm} (in Appendix), gives the desired result. Note that $A(t)$ being Hurwitz and row-diagonally dominant implies that $\alpha_r(t')<0$.
\qed

In particular, 
consider a time-varying system $\dot x = -A(t) x$ with $A(t) : = \{a_{ij}(t)\} \in \mathbb{R}^{n \times n}$ satisfying
 \begin{align} \label{eq:column_dDD}
    a_{ii}(t) - \sum_{j =1, i\neq j}^n |a_{ij}(t)| \geq \delta >0, \forall i = 1,2,\cdots, n; \forall t \geq \bar t
\end{align}
with a finite time $\bar t$.  Then all of its solutions converge to zero exponentially fast with the rate $e^{-\delta t}$ as $t \rightarrow \infty$. In fact, it holds that
\begin{align} \label{eq:exponential_bound1}
   |x_i(t)| \leq  \|x(t)\|_\infty \leq  \|x(\bar t)\|_\infty e^{-\delta t}, \forall i, \forall t \geq \bar t.  
\end{align}

\begin{lemma} \label{lemma:column_exponential}
(Column-diagonally dominant linear system)
Consider a time-varying linear system $\dot x(t) = A(t)x(t)$, where $A(t)$ is a continuous-time Hurwitz matrix with column-diagonally dominant entries $\forall t \geq t_0$. Then it holds that
\begin{align}
        \|x(t)\|_1 \leq \|x(t_0)\|_1 e^{\int_{t_0}^{t} \alpha_c(t') \text{d}t'}, \forall t \geq t_0, 
\end{align}
where $\alpha_c(t') = \text{max}_{j =1, 2, \cdots, n} \left(a_{jj}(t') +\sum_{i=1, i\neq j}^n |a_{ij}(t')|\right)$ and $\alpha_c(t')<0$.
\end{lemma}
\textbf{Proof}
Applying Lemma~\ref{theorem:measure_exponential}, and choosing the vector norm as the one-norm with the matrix measure induced by the vector one-norm in \eqref{eq:measure_one_norm} (in Appendix), gives the desired result. Note that $A(t)$ being Hurwitz and column-diagonally dominant implies that $\alpha_c(t')<0$.
\qed

In particular, 
consider a time-varying system $\dot x = -A(t) x$ with $A(t) : = \{a_{ij}(t)\} \in \mathbb{R}^{n \times n}$ satisfying
 \begin{align} \label{eq:column_dDD}
    a_{jj}(t) - \sum_{i =1, i\neq j}^n |a_{ij}(t)| \geq \delta >0, \forall j = 1,2,\cdots, n, \forall t \geq \bar t 
\end{align}
with a finite time $\bar t$. Then all of its solutions converge to zero exponentially fast with the lower bound rate $e^{-\delta t}$ as $t \rightarrow \infty$. In fact, it holds that
\begin{align} \label{eq:exponential_bound1}
   |x_i(t)| \leq  \sum_{i = 1}^n |x_i(t)| \leq e^{-\delta t} \sum_{i = 1}^n |x_i(\bar t)|, \forall i,  \forall t \geq \bar t.   
\end{align}
We note that the solution bound and exponential convergence in \eqref{eq:exponential_bound1} under the condition of \eqref{eq:column_dDD} for column-diagonally  dominant  linear  systems generalize the main Theorem of \cite{kahane1972stability}.

\section{Distributed high-gain adaptive stabilization: System (I) case} \label{sec:main_matrix_gain}
\subsection{Matrix high-gain stabilizability} \label{sec:stabilizable_matrix_gain}

Consider the following uncertain closed-loop linear system 
\begin{align} \label{eq:system}
    \dot x(t) = (A - K(t)B)x(t), t\geq 0,
\end{align}
where $x(t) \in \mathbb{R}^n$ is the system state, $A \in \mathbb{R}^{n\times n}$, $B \in \mathbb{R}^{n\times n}$ are system matrices, and $K(t) : = \text{diag}\{k_1(t), k_2(t), \allowbreak \cdots, k_n(t)\} \in \mathbb{R}^{n\times n}$ is a positive diagonal gain matrix  with each time-varying individual gain  $k_i(t)$ being positive and monotonically increasing.   We remark that the system matrices $A$ and/or $B$ are \textbf{unknown}, and we aim to design adaptive control laws that regulate the gain matrix $K(t)$ such that the uncertain system \eqref{eq:system} is stabilized. The adaptive gain matrix $K(t)$ renders that the system \eqref{eq:system} is a time-varying linear  system. Therefore, the system matrix $(A - K(t)B)$ being Hurwitz  at all the time does \textit{not} necessarily conclude convergence, and additional conditions are required to ensure system stability \cite{rugh1996linear}.

The definition of matrix high-gain stabilizability under a diagonal gain matrix is given below.

\begin{definition}
(Matrix high-gain stabilizability) A multivariable linear system \eqref{eq:system} (or \eqref{eq:system_example_II}) is stabilizable by  high-gain diagonal matrix functions $K(t)$ if there exist  a positive constant  $\bar k$ and a finite time $\bar t$, such that for $k_i(t)>\bar k, \forall i, \forall t>\bar t$ the system  is asymptotically stable with the diagonal matrix  $K(t)$ and its solutions $x(t)$ converge to zero exponentially fast.
\end{definition}

The following example shows the distinct features between scalar high-gain stabilizability (the Byrnes-Willems condition \cite{byrnes1984adaptive}) and matrix high-gain stabilizability. 
\begin{example}
Consider a second-order  uncertain system with unknown system matrices $A, B \in \mathbb{R}^{2 \times 2}$, while the true value of the  matrix $B$ is given by 
\begin{align} \label{eq:matrix_example}
B = \left[
\begin{array}{cc}
  2 &3  \\ -1 & -1   
    \end{array} \right].
\end{align}
Clearly $-B$ is a Hurwitz matrix with eigenvalues $\lambda_{1,2}(-B) = -0.5000 \pm 0.8660i$. The matrix $-k(t)B$ with any positive scalar function $k(t)>0$ remains a Hurwitz matrix, and   by the  Byrnes-Willems Theorem \cite{byrnes1984adaptive} the uncertain linear system is stabilizable by a scalar high gain $k(t) >\bar k$ where $\bar k$ is a sufficiently large threshold gain value. 

However, we show that a linear system with the  matrix $B$ in \eqref{eq:matrix_example} is not stabilizable by diagonal matrix high gains $K(t) = \text{diag}\{k_1(t), k_2(t)\}$. Without loss of generality we assume $A = 0$ and consider the   system $\dot x(t) = -K(t)B x(t)$. The matrix measure of $K(t)B$, with the induced matrix norm  chosen by the column-sum norm, is calculated by $\mu(K(t)B) = \text{max}(2k_1(t) - k_2(t), 3k_1(t)-k_2(t))$. For $k_2(t)>3k_1(t)$, we have $\mu(K(t)B) = 3k_1(t) - k_2(t) <0$, and by Lemma~\ref{theorem:measure_exponential}, one concludes that the solution satisfies 
$
    \|x(t)\|_1 \geq \|x(0)\|_1 e^{- \int_{t_0}^{t} \mu(K(t')B)\text{d}t'}  
    = \|x(0)\|_1 e^{\int_{t_0}^{t} (k_2(t')- 3k_1(t')))\text{d}t'}
$
     which grows unbounded for $k_2(t)>3k_1(t)$.   Therefore, the system with a Hurwitz matrix $-B$ given in \eqref{eq:matrix_example} is not stabilizable by matrix high gains no matter how large one chooses $k_1(t)$ and $k_2(t)$ in $K(t)$  (In this example, the stability indeed depends on the \textit{relative} magnitude of each individual gain $k_i(t)$ in the matrix gain function $K(t)$).  
\end{example}

In this section, 
we first give a characterization of uncertain multivariable systems that are stabilizable by matrix high gains, and then provide a state-dependent updating law for adaptively adjusting matrix gains to ensure system convergence with upper bounded and convergent matrix gains. It turns out that H-matrices and \textit{generalized diagonally dominant systems} \cite{willems1976lyapunov} play an important role in tuning adaptive   matrix gains for stabilizing uncertain multivariable linear systems. 
The intuition is that with the increasing of each diagonal entry of $K(t)$, the term $-K(t)B$ should dominate the unknown matrix $A$ under each sufficiently large $k_i(t)$. 

The following theorem   characterizes an important set of   uncertain linear systems that are stabilizable by matrix high gains. This can be   seen as a counterpart to  the classical results on uncertain linear systems stabilizable by \textit{scalar} high gains (see e.g., \cite[Theorem 3.5]{maartensson1986adaptive} and \cite[Proposition 2.1]{ilchmann1987high}).

\begin{theorem} \label{theorem:infinite_gain}
(High gain stabilizability) Consider the uncertain linear system \eqref{eq:system} with unknown system matrices $A$ and/or $B$.  Suppose $B$ is an \textbf{$H$-matrix with   positive diagonal entries}, and each individual gain function $k_i(t)$ in the matrix gain $K(t)$ is a monotonically increasing function approaching infinity as $t \rightarrow \infty$. Then the uncertain linear system \eqref{eq:system} is exponentially convergent to zero. 
\end{theorem}

\textbf{Proof}
From Theorem~\ref{theorem:_H_matrix}, the condition that the matrix $B$ is an H-matrix implies that there exists a positive   diagonal matrix $\bar D = \text{diag}\{\bar d_1, \bar d_2, \cdots, \bar d_n\}$ such that $\bar B = \{\bar b_{ij}\} := \bar D^{-1}B \bar D$ is  strictly \textbf{row-diagonally} dominant. 
  By a coordinate transform $z : = \bar D^{-1} x$ we will consider the system 
\begin{align} \label{eq:z_system_row_dd}
    \dot z &= \bar D^{-1} \dot x = \bar D^{-1}(A - K(t)B)\bar D z \nonumber \\
     &= \left(\bar D^{-1}A\bar D - K(t)\bar D^{-1}B\bar D \right) z.
\end{align}
Note that the commuting equality $\bar D^{-1}K(t) = K(t)\bar D^{-1}$ holds since both matrices $\bar D^{-1}$ and $K(t)$ are diagonal.  
Note also that the diagonal entries of $\bar D^{-1}B\bar D$ satisfy $\bar b_{ii} = b_{ii}, \forall i$, which are positive. Let $\bar a_{ij}$ denote  the $(ij)$-th entry of the matrix $\bar D^{-1}A\bar D$. Due to the strict row-diagonal  dominance of the matrix $\bar B$, it holds that   $\bar b_{ii} - \sum_{j=1, j\neq i}|\bar b_{ij}| >0, \forall i$. Now by choosing $\bar k_i$ such that  
\begin{align}
     k_i(t) \geq \bar k_i  = \frac{\sum_{j=1, j\neq i}  |\bar a_{ij}| + \bar a_{ii} +\delta}{\bar b_{ii} - \sum_{j=1, j\neq i}|\bar b_{ij}|}, 
\end{align}
where $\delta >0$ is any positive constant predefined, it holds that
\begin{align} \label{eq:diagonal_inequality_row}
     k_i(t)\bar b_{ii} - \bar a_{ii}  - \left(\sum_{j=1, j\neq i}(  |\bar a_{ij}| + k_i(t)|\bar b_{ij}| )  \right) \geq \delta >0.
\end{align}
Note that since all entries of $A$ and $\bar B$ are bounded, all $\bar k_i, \forall i$ are bounded. 
Choose $\bar k := \text{max}_i \bar k_i$ and let $k_i(t) \geq \bar k, \forall i, \forall t>\bar t$. Then \eqref{eq:diagonal_inequality_row} holds $\forall i = 1,2,\cdots, n, \forall t>\bar t$. 
By Lemma~\ref{lemma:row_dd_system} this proves that the $z$ system \eqref{eq:z_system_row_dd} converges to zero exponentially fast with a least convergence rate $e^{-\delta (t - \bar t)}$, $\forall t >\bar t$. This in turn implies that the linear system \eqref{eq:system} converges to zero exponentially fast with the convergence scaled by the coordinate transform  $x := \bar D z$.  
 \qed

\begin{remark}  \label{remark:for_theorem2}
Some remarks of Theorem~\ref{theorem:infinite_gain} are in order. 
\begin{itemize}
    \item The condition $k_i(t) \rightarrow \infty$ as $t \rightarrow \infty$ is not really used in the proof.   So long as the condition in \eqref{eq:diagonal_inequality_row} is satisfied that guarantees $k_i(t) \geq \bar k, \forall i, \forall t>\bar t$, the linear system \eqref{eq:z_system_row_dd} is exponentially convergent with a rate $e^{-\delta (t - \bar t)}$, $\forall t >\bar t$. Nevertheless,  we follow the same spirit of scalar high-gain stabilizability (\cite[Theorem 3.5]{maartensson1986adaptive} and \cite[Proposition 2.1]{ilchmann1987high}) to state the matrix high-gain stabilizability in Theorem~\ref{theorem:infinite_gain}. 
    \item  With the condition  $k_i(t) \rightarrow \infty$ as $t \rightarrow \infty$, one can claim a stronger result termed   \textit{arbitrarily fast exponential convergence} \cite{ilchmann1987high}; i.e., the exponential rate $\delta(t)$ is a monotonically increasing function of the time $\forall t> \bar t$, and $\delta(t) \rightarrow \infty$ as $t \rightarrow \infty$. This is also evident   by   Lemma~\ref{lemma:row_dd_system}. 
    \item The sufficient condition for the uncertain system \eqref{eq:system} being matrix high-gain stabilizable is that,   with a sufficiently large diagonal matrix gain $K(t)$, the system \eqref{eq:system} should become a generalized  diagonally row dominant  system at some finite time $\bar t$ and will remain it $\forall t>\bar t$ so as to ensure the exponential stability of the system states.  Note that the exponential rate also grows with the growing $k_i(t)$, for $t> \bar t$.
\end{itemize}

\end{remark}

\subsection{Matrix high-gain updating laws for distributed  adaptive stabilization}
 
Now we are ready to show one of the main results of this paper. 
The following theorem can be seen as the matrix high-gain extension of the classical Byrnes-Willems Theorem \cite{byrnes1984adaptive} on adaptive scalar high-gain stabilization.

\begin{theorem} \label{theorem:main_gain_updating}
Consider the uncertain linear multivariable system \eqref{eq:system} with unknown system matrices $A$ and/or $B$, and suppose that  $B$ is an H-matrix with   positive diagonal entries. Each individual gain $k_i(t)$ in the adaptive matrix gain $K(t) = \text{diag}\{k_1(t), k_2(t), \cdots, k_n(t)\}$ is updated by the following distributed adaptive law
\begin{align} \label{eq:updating_ki}
    \dot k_i(t) = c_i|x_i(t)|^{p_i}, \,\,   k_{i}(0)>0, 
\end{align}
where $c_i,  p_i$ are positive constants with $c_i>0, p_i \geq 1$. Then the following statements hold.
\begin{enumerate}[(i)]
    \item The solutions to the linear system \eqref{eq:system} and  the adaptive gain updating system \eqref{eq:updating_ki} always exist, are unique, and can be extended to $t \rightarrow \infty$.
    \item The uncertain system \eqref{eq:system} with unknown system matrices $A, B$ is stabilized with the adaptive matrix gain $K(t)$ in the sense that $\text{lim}_{t \rightarrow \infty} x(t) = 0$.
    \item Each distributed gain $k_i(t)$ in the adaptive matrix gain $K(t)$ is monotonically increasing, upper bounded and convergent in the limit in the sense that $\text{lim}_{t \rightarrow \infty} k_i(t) = k^i_{\infty} <\infty, \forall i$, where $k^i_{\infty}$ is a bounded positive constant. 
\end{enumerate}
\end{theorem}
\textbf{Proof}
For the time-varying linear system \eqref{eq:system}, the existence and uniqueness of the solution can be ensured if the state matrix $(A - K(t)B)$ is continuous and uniformly bounded (see e.g., \cite[Chapter 1.2]{brockett1970finite}), which is equivalent to that the gain matrix $K(t)$ is continuous and uniformly bounded. Since each diagonal entry $k_i(t)$ of the gain matrix $K(t)$ is updated by the differential equation \eqref{eq:updating_ki}, the gain matrix $K(t)$ is a differentiable function of time and thus is uniformly continuous. Therefore the solutions to the linear system \eqref{eq:system} and  the adaptive gain updating system \eqref{eq:updating_ki} uniquely  exist. 
Since  each $k_i(t)$ cannot increase to be unbounded at any finite time, the solutions can be extended to $t \rightarrow \infty$. In the following analysis we will rule out the possibility that $x(t)$ or $k_i(t)$ becomes unbounded in the limit $t \rightarrow \infty$.

 According to the updating law \eqref{eq:updating_ki}, each distributed gain function $k_i(t)$ will keep increasing as long as  $|x_i(t)| \neq 0$. Since the unknown matrix $A$ is bounded   there must exist a finite time $\bar t$ such that the condition in Eq.~\eqref{eq:diagonal_inequality_row} of Theorem~\ref{theorem:infinite_gain} holds, \footnote{The existence of such a finite time $\bar t$ can be shown by contradictions.  Suppose no such finite time $\bar t$ exists, which then implies the system state $x_i(t)$ is unstable and is non-zero for almost all the time. By integrating \eqref{eq:updating_ki} this in turn implies that $k_i(t)$ will always keep increasing, while an unbounded $k_i(t)$   leads to an arbitrarily fast exponential convergence of all system states according to Theorem~\ref{theorem:infinite_gain}, thus a contradiction. For adaptive stabilization of scalar uncertain linear systems, the existence of such a finite time $\bar t$ is discussed in \cite{Ilchmann}. } and therefore from Lemma~\ref{lemma:row_dd_system} one can show 
\begin{align} \label{eq:exponential_t*}
|x_i(t)| \leq M e^{-\delta(t - \bar t)}, \forall t\in [\bar t, \infty),
\end{align}
with some positive constants $M>0$ and $\delta>0$. \footnote{ According to Lemma~\ref{lemma:row_dd_system} and the convergence inequality \eqref{eq:exponential_bound1}, the constant $M$ is related to a state norm at the time $\bar t$, and the exponential decay rate $\delta$ will continue to increase with the monotonic increasing of each individual gain $k_i(t)$  for $t \geq \bar t$. }
This implies that the state $x(t)$ must converge to the origin exponentially fast $\forall t\in [\bar t, \infty)$. 

Now we prove the third statement. Note from the existence of a finite time $\bar t$ as ensured in Theorem~\ref{theorem:infinite_gain} and the exponential convergence inequality  \eqref{eq:exponential_t*}, one has 
\begin{align} \label{eq:converge_ki}
    k_i(\infty) &= k_i(0)+ \int_0^\infty {c_i|x_i(t)|^{p_i}} \text{d} t  \nonumber \\
    &= k_i(0)+ c_i \left(\int_0^{\bar t} {|x(t)|^{p_i}} \text{d} t + \int_{\bar t}^\infty {|x(t)|^{p_i}} \text{d} t \right) \nonumber \\
    & \leq k_i(0)+  c_i \left(\int_0^{\bar t} {|x(t)|^{p_i}} \text{d} t + \frac{M}{\delta p_i} \right).
\end{align}
Since $|x_i(t)|$ is bounded at the finite time interval $[0, \bar t)$, the first integral $\int_0^{\bar t} {|x_i(t)|^{p_i}} \text{d} t$ is bounded; furthermore, due to the exponential convergence inequality in \eqref{eq:exponential_t*}, the second integral $\int_{\bar t}^\infty {|x_i(t)|^{p_i}} \text{d} t$ is also upper bounded by $ \frac{M}{\delta p_i}$. Therefore, $k_i(t)$ is upper bounded in the limit. Also note that each $k_i(t)$ is continuous and monotonically increasing; therefore it must converge to some bounded value $k^i_{\infty}$. \footnote{This is due to the well-known result: if a continuous  function $f(t):  [a, \infty) \rightarrow \mathbb{R}$ is increasing and bounded from above, then $\text{lim}_{t \rightarrow \infty} f(t)$ exists and is finite.} This completes the proof.  
\qed

\begin{remark}
We note that the overall stabilization control system consisting of the uncertain linear system \eqref{eq:system} and  the adaptive gain updating system \eqref{eq:updating_ki} is nonlinear,  due to the nonlinearity   in the updating law \eqref{eq:updating_ki}. Therefore, in general it is hard or even impossible to give an analytical formula for the converged
values of each distributed gain.  However,   the inequality in \eqref{eq:converge_ki} shows an estimate of the upper bound of each individual gain $k_i(t)$. Clearly,  with larger values of $c_i$ and $k_i(0)$, the upper bounds of $k_i(\infty)$ will be larger. Furthermore, the positive parameters $c_i,  p_i$ can be used to adjust the growing speed of each updating function: under transient system state $|x_i(t)| \geq 1$, larger values of $c_i,  p_i$ lead to a larger updating function $c_i|x_i(t)|^{p_i}$, and thus the updating and growing speed for $k_i(t)$ is  increased. As a consequence, the finite time $\bar t$ when the uncertain system  starts to exponentially decay can be shortened. 
\end{remark}

\begin{remark}
Under the structural assumption  on the unknown matrix $B$, this
non-linear (adaptive) controller \eqref{eq:updating_ki} is robust to large structural uncertainties  for the uncertain linear system \eqref{eq:system} (apart from the boundedness condition,  no other condition on $A$ is required). In contrast, a linear
non-adaptive controller \cite{green2012linear} would give very limited robustness to structural 
uncertainty in $A$ and $B$. We also remark that the proposed distributed adaptive law inherits similar robustness properties from the scalar gain adaptive stabilization law (e.g., \cite{ilchmann1987high}). To further improve the robustness property of  uncertain control systems with unstructured 
uncertainties, external or non-vanishing  perturbations,  the robust adaptive techniques (such as the $\sigma$ modification, dead zone approach, dynamic projection method, etc.) discussed  in \cite{ioannou2012robust} can be adopted to design robust distributed adaptive law.
\end{remark}

\subsection{Numerical examples} 
We show some numerical simulation examples to illustrate the main results of this section.   Consider an uncertain linear system \eqref{eq:system} with unknown system matrices $A$ and $B$, while for simulation purpose  their true values are chosen as
\begin{align} \label{eq:simulation_matrix}
A = \left[
\begin{array}{ccc}
  1 &4 &2 \\ 5 & -2 & 1 \\ 6 & 3 & -4 
    \end{array} \right],
B = \left[
\begin{array}{ccc}
7 &4 &-2\\ -4 &6 &3 \\ 2 &-2& 5
    \end{array} \right]                   .
\end{align}
In this simulation example it can be verified that $B$ is an H-matrix (its comparison matrix $M_B$, which has eigenvalues $\lambda_{1,2,3}(B) =  \{0.3184, 10.5111, 7.1705\}$, is an M-matrix). Note the matrix $B$ in \eqref{eq:simulation_matrix} is not diagonally dominant by itself, but  can be made generalized diagonally dominant by some positive diagonal matrix according to Theorem~\ref{theorem:_H_matrix}.   Theorems~\ref{theorem:infinite_gain} and~\ref{theorem:main_gain_updating} indicate that the uncertain linear system \eqref{eq:system} with the unknown system matrices in \eqref{eq:simulation_matrix} is stabilizable by the adaptive matrix high gain $K(t)$ updated by the distributed state-dependent adaptive law \eqref{eq:updating_ki}. 

\begin{figure}[t]
\begin{center}
\includegraphics[width=0.4\textwidth]{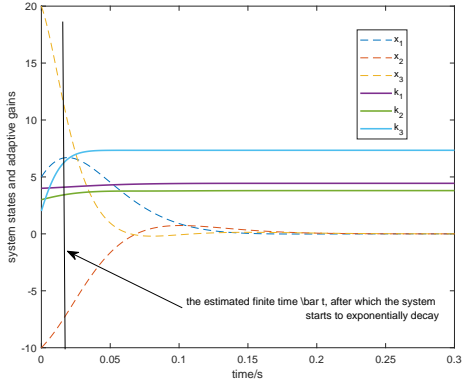}
\caption{Adaptive stabilization of an uncertain system \eqref{eq:system} with distributed adaptive matrix gains. The parameters in the updating functions   are set as $c_i = 1, i = 1, 2,3$ and $p_1 = 1, p_2 = 1.5, p_3 = 2$. }
\label{fig:System_1_examp1}
\end{center}
\end{figure}

\begin{figure}[t]
\begin{center}
\includegraphics[width=0.4\textwidth]{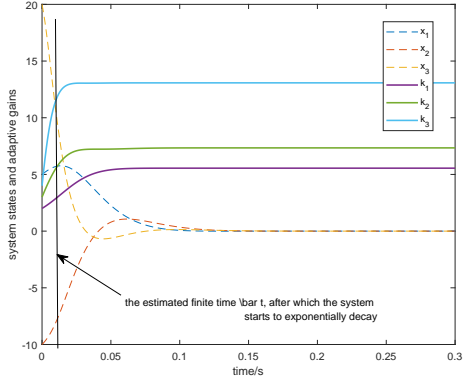}
\caption{Adaptive stabilization of an uncertain system \eqref{eq:system} with distributed adaptive matrix gains. The parameters in the updating functions   are set as $c_i = 3, i = 1, 2,3$ and $p_1 = 2, p_2 = 2, p_3 = 2$.  }
\label{fig:System_1_examp2}
\end{center}
\end{figure}

In the simulations, the initial values for the system states are chosen by $x(0) = [5, -10, 20]^T$, and the initial values of adaptive gains are set  as $k_1(0) = 4, k_2(0) = 3$ and $k_3(0) = 2$.   The simulation results that demonstrate  convergences of both system states and distributed adaptive matrix gains are shown in Fig.~\ref{fig:System_1_examp1} and Fig.~\ref{fig:System_1_examp2} under different values of the updating function parameters $c_i$ and $p_i$. Clearly, without identifying the true values of the unknown matrices $A$ and $B$,  the adaptive matrix gains updated by \eqref{eq:updating_ki} guarantee that the system states converge to zero exponentially fast, while all distributed adaptive gains monotonically converge to some constant and bounded values.  
Furthermore,  it can be clearly observed in Fig.~\ref{fig:System_1_examp1} and Fig.~\ref{fig:System_1_examp2} that the updating speed for the three distributed gains $k_i(t)$ is increased with larger values $c_i$ and $p_i$. As a consequence, the finite time $\bar t$ when the uncertain system  starts to  exponentially decay   has also been shortened, which implies the uncertain system settles down more rapidly by a faster updating of each individual gain.

\section{Distributed high-gain adaptive stabilization: System (II) case} \label{sec:main_matrix_gain_II}
\subsection{Matrix high-gain stabilizability} \label{sec:stabilizable_matrix_gain_II}

In this section we provide some corresponding results  for  uncertain linear systems in the form of System (II) in \eqref{eq:system_example_II}. The main analysis follows from the equivalence results in Theorem~\ref{theorem:_H_matrix}.

\begin{theorem} \label{theorem:infinite_gain_BK}
(High-gain stabilizability) Consider the uncertain linear system  
\begin{align} \label{eq:system_BK}
    \dot x(t) = (A - BK(t))x(t), t\geq 0,
\end{align}
where the system matrices $A$ and/or $B$ are unknown.  Suppose $B$ is an \textbf{$H$-matrix with   positive diagonal entries},  and each gain entry $k_i(t)$ in the matrix gain $K(t)$ is a monotonically increasing function approaching infinity as $t \rightarrow \infty$. Then the uncertain linear system \eqref{eq:system_BK} is exponentially convergent to zero. 
\end{theorem}

\textbf{Proof}
The proof follows a similar spirit as that of Theorem~\ref{theorem:infinite_gain}, but we will focus on the column-diagonal dominance of the matrix $B$. From Theorem~\ref{theorem:_H_matrix}, the matrix $B$ being an H-matrix implies that there exists a positive   diagonal matrix $\tilde D = \text{diag}\{\tilde d_1, \tilde d_2, \cdots, \tilde d_n\}$ such that $\tilde B = \{\tilde b_{ij}\} :=\tilde D B \tilde D^{-1}$ is  strictly  column-diagonally dominant. 
  By a coordinate transform $z : = \tilde D x$ we will consider the transformed system 
\begin{align} \label{eq:z_system}
    \dot z &= \tilde D \dot x = \tilde D(A - BK(t))\tilde D^{-1} z \nonumber \\
     &= \left(\tilde DA\tilde D^{-1} - \tilde DB\tilde D^{-1}K(t) \right) z. 
\end{align}
Note  that the diagonal entries of $\tilde DB\tilde D^{-1}$ satisfy $\tilde b_{ii} = b_{ii}, \forall i$, which are positive.
Let $\tilde a_{ij}$ denote  the $(ij)$-th entry of the matrix $\tilde DA\tilde D^{-1}$. Due to   the strict column diagonal  dominance of the matrix $\tilde B$ it holds that   $\tilde b_{jj} - \sum_{i=1, i\neq j}|\tilde b_{ij}| >0, \forall j$. Now by choosing $\tilde k_j$ such that  
\begin{align}
     k_j(t) \geq \tilde k_j  = \frac{\sum_{i=1, i\neq j}  |\tilde a_{ij}| + \tilde a_{jj} +\delta}{\tilde b_{jj} - \sum_{i=1, i\neq j}|\tilde b_{ij}|}, 
\end{align}
where $\delta >0$ is any positive constant predefined, it holds that
\begin{align} \label{eq:diagonal_inequality}
     k_j(t)\tilde b_{jj} - \tilde a_{jj}  - \left(\sum_{i=1, i\neq j}(  |\tilde a_{ij}| + k_j(t)|\tilde b_{ij}| )  \right) \geq \delta >0.
\end{align}

Note that since all entries of $A$ and $\tilde B$ are bounded, all $\tilde k_j, \forall j$ are also bounded.  Choose $\tilde k := \text{max}_j \tilde k_j$ and let $k_j(t) \geq \tilde k, \forall j, \forall t>\bar t$. Then \eqref{eq:diagonal_inequality} holds $\forall j = 1,2,\cdots, n, \forall t>\bar t$.  
By Lemma~\ref{lemma:column_exponential} this proves that the $z$ system \eqref{eq:z_system} converges to zero exponentially fast with the least convergence rate $e^{-\delta (t - \bar t)}$, $\forall t >\bar t$. This in turn implies that the linear system \eqref{eq:system_BK} converges to zero exponentially fast with the convergence scaled by the coordinate transform  $x :=\tilde D^{-1} z$.   
\qed

Again, we remark that all statements in Remark~\ref{remark:for_theorem2} also apply to Theorem~\ref{theorem:infinite_gain_BK}.

\subsection{Matrix high-gain updating laws for distributed adaptive stabilization}
The corresponding result on distributed gain updating law and system convergence for  System (II) is shown below. 
\begin{theorem} \label{theorem:main_gain_updating_BK}
Consider the uncertain linear multivariable system \eqref{eq:system_BK} with unknown system matrices $A$ and/or  $B$. Suppose that the matrix $B$ is an \textbf{$H$-matrix with   positive diagonal entries}. Each individual gain $k_i(t)$ in the adaptive matrix gain $K(t) = \text{diag}\{k_1(t), k_2(t), \cdots, k_n(t)\}$ is updated by the following distributed adaptive law
\begin{align} \label{eq:updating_ki_BK}
    \dot k_i(t) = c_i|x_i(t)|^{p_i},     k_{i}(0)>0, 
\end{align}
where $c_i,  p_i$ are positive constants with $c_i>0, p_i \geq 1$. Then the following statements hold.
\begin{enumerate}[(i)]
    \item The solutions to the linear system \eqref{eq:system_BK} and  the adaptive gain updating system \eqref{eq:updating_ki_BK} always exist, are unique, and can be extended to $t \rightarrow \infty$.
    \item The uncertain system \eqref{eq:system_BK} with unknown system matrices $A, B$ is stabilized with the adaptive matrix gain $K(t)$ in the sense that $\text{lim}_{t \rightarrow \infty} x(t) = 0$.
    \item Each distributed gain $k_i(t)$ in the adaptive matrix gain $K(t)$ is monotonically increasing, upper bounded and convergent in the limit in the sense that $\text{lim}_{t \rightarrow \infty} k_i(t) = k^i_{\infty} <\infty, \forall i$, where $k^i_{\infty}$ is a bounded positive constant. 
\end{enumerate}
\end{theorem}
\textbf{Proof}
The proof is based on the result of Theorem~\ref{theorem:infinite_gain_BK}, and follows similar steps as that of Theorem~\ref{theorem:main_gain_updating}. It is omitted here for brevity.
\qed

\subsection{Numerical examples}
We consider an uncertain control system \eqref{eq:system_BK} in $\mathbb{R}^5$ with unknown system matrices, whose true values are given by 
\begin{align} \label{eq:matrix_B}
    A =  \left[ \begin{array}{ccccc}
    1.9790  &  0.5275  &  1.7078  &  0.8509  &  0.5712 \\
    1.2670  &  1.8672  &  1.3512  &  1.2259  &  1.2712 \\
    1.4203  &  1.9343  &  1.2113  &  0.7842  &  0.3338 \\
    0.7864  &  1.8849  &  0.2922  &  1.1515  &  0.5984 \\
    0.2575  &  1.9451  &  0.2046  &  1.7284  &  0.9200  
    \end{array} \right],
\end{align}
and
\begin{align} \label{eq:matrix_B}
    B =  \left[ \begin{array}{ccccc}
    1.1837 &   0.1407   & 0.2400  & -0.4138   & 0.2894 \\
   -0.3679  &  1.3288  & -0.2652 &  -0.1336 &  -0.1323 \\
    0.2227  &  0.1538  &  0.7350  & -0.1308   &-0.2940 \\
   -0.3896  &  0.2491  &  0.4706  &  1.1850 &  -0.4133 \\
   -0.3825  &  0.0832   & 0.3669  &  0.0979  &  1.1719 
    \end{array} \right].
\end{align}
The two matrices are generated randomly in Matlab for the simulation purpose. It is verified that the comparison matrix of $B$, denoted by $M_B$, is an M-matrix (whose eigenvalues are $\lambda(M_B) = [0.0456, 1.0995,  1.547 4 + 0.1286i, 1.5474 - 0.1286i, 1.3645]$) and therefore the matrix $B$ is an H-matrix. 
In the numerical simulation we set the initial conditions for the states and gain matrix as $x(0) = [10, -10, -15, 15, -8]^T$ and $K(0) = \text{diag}\{2,  3,  4, 5, 6\}$. The updating functions for distributed adaptive gains are chosen as $f_i  = |x_i(t)|^{1.5}$, i.e., $c_i = 1, p_i = 1.5, \forall i$. 

\begin{figure}[t]
\begin{center}
\includegraphics[width=0.55\textwidth]{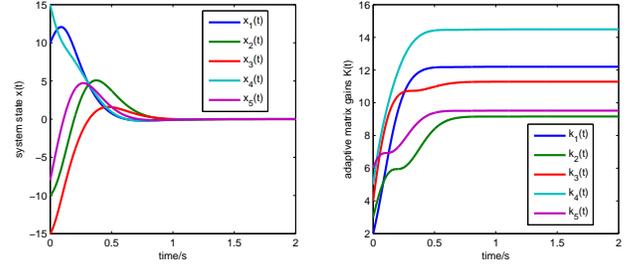}
\caption{Adaptive stabilization of an uncertain system \eqref{eq:system_BK} with distributed adaptive matrix gains. Left: convergence of system states. Right: convergence of distributed adaptive gains. }
\label{fig:Simu_2_BK}
\end{center}
\end{figure}

The simulation results that demonstrate  convergences of both system states and distributed adaptive gains are shown in Fig.~\ref{fig:Simu_2_BK}. Clearly, without identifying the true values of the unknown matrices $A$ and $B$, distributed adaptive   gains updated by \eqref{eq:updating_ki_BK} guarantee that the system states converge to zero exponentially fast, while all individual adaptive gains monotonically converge to some constant and bounded values. In this simulation, 
we observe that the final converged values for each individual gain are $k_1(\infty)\approx12.2056,  k_2(\infty)\approx9.1612$,  $k_3(\infty)\approx 11.2881$, $k_4(\infty)\approx14.4884$,  and  $k_5(\infty)\approx9.5236$, as shown in the right figure of Fig.~\ref{fig:Simu_2_BK}.

\begin{figure}[htb]
 \begin{center}
  \includegraphics[width=0.8\linewidth]{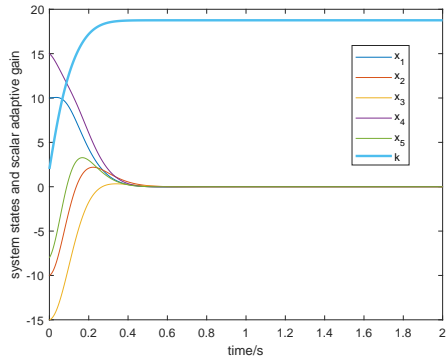} \end{center}
 \caption{  Adaptive stabilization of uncertain multivariable systems, by a scalar adaptive gain: the system is described by $\dot x(t) = (A - k(t)B) x(t)$, with a scalar adaptive gain $k(t)$. }
 \label{fig:example_II_scalar}
\end{figure}

For a comparative study, we also simulate the stabilization control of the above uncertain linear system with an adaptive scalar gain $k(t)$ updated by all state information.  The scalar gain $k(t)$ is updated by the adaptive law $\dot k(t) = \|x(t)\|^{1.5}$, with an initial condition $k(0) = 2$. In Fig.~\ref{fig:example_II_scalar}, one can observe that the scalar gain $k(t)$ grows unnecessarily larger than  the distributed matrix gains  in Fig.~\ref{fig:Simu_2_BK}.  Furthermore, the updating of the scalar gain $k(t)$ involves all five system states, which will soon become impractical when the uncertain system includes a large number of states or some state information is inaccessible. In contrast, the distributed adaptive stabilization scheme offers several  
advantages   such as   low computational complexity, improved scalability and high 
flexibility. These advantages are significant in the control task involving large-scale systems, as will be discussed in the next section on adaptive synchronization of complex networks.

\section{Applications to distributed adaptive synchronization of complex networks} \label{sec:network_synchronization}
In this section we discuss a typical application of  distributed adaptive stabilization theory in distributed and scalable control of large-scale networked systems. The application example  involves distributed adaptive synchronization of complex networks, in which the coupling weights are adaptively adjusted by local information to reach network synchronization. 

Following \cite{wu1995synchronization, nijmeijer1997observer, dorfler2014synchronization}
we consider the following   complex network system
\begin{align} \label{eq:node_network_dynamics}
    \dot x_i(t) = f(x_i(t)) + k_i(t) \sum_{j \in \mathcal{N}_i} (x_j(t) - x_i(t)),
\end{align}
where $x_i \in \mathbb{R}^n$ is the  system state of the $i$-th node, $f(\cdot)$ is the system vector function, $\mathcal{N}_i$ denotes the neighboring set for node $i$, and $k_i >0$ is a local coupling gain (or coupling weight) for node $i$ associated with the diffusive coupling $x_j - x_i$. Synchronization control for complex network systems aims to achieve $x_i(t) \rightarrow x_j(t)$, $\forall i,j$, with $t \rightarrow \infty$. 
  The condition on reaching network synchronization  depends on the (possibly unknown) vector function $f$,  the coupling weight $k_i$, and the graph connection, which often involves the network connectivity and topology. In practice, it is prohibitive to derive such synchronization conditions for large-scale networks, since the unknown dynamics functions are hard to estimate, and the network connectivity (in terms of Laplacian spectrum) involves global network information and its calculation is computationally expensive.   Adaptive synchronization with distributed adaptive time-varying coupling weights $k_i(t)$ is preferable in particular for large-scale networks, since it avoids using any global information for achieving network synchronization even with unknown system dynamics and little knowledge of network topology. 

Before presenting the main result we impose the following condition on the vector function $f$, which is a standard assumption commonly used in the study of complex network synchronization (see e.g.,  \cite{chen2007pinning,delellis2011quad, montenbruck2015practical}). 
\begin{definition} \label{Def:QUAD_condition}
A function $f:\mathbb{R}^{n} \times {\mathbb R}^{+} \mapsto {\mathbb R}^{n}$ is said to be QUAD $(\Delta, \epsilon)$ if and only if, for any $x$, $y\in{\mathbb R}^{n}$, it holds that
\begin{align} \label{eq:QUAD}
&(x-y)^{T}[f(x,t)-f(y,t)]-(x-y)^{T}\Delta (x-y)  \nonumber \\
&\leq -\epsilon (x-y)^{T}(x-y),
\end{align}
where $\Delta$ is an $n \times n$ diagonal matrix and $\epsilon$ is a real  scalar.
\end{definition}

By defining $x = [x_1^T, x_2^T, \cdots, x_N^T]^T, f(x) = [f(x_1)^T, f(x_2)^T, \allowbreak \cdots, f(x_N)^T]^T$, the diagonal coupling weight matrix $K(t) = \text{diag}\{k_1(t), k_2(t), \cdots, k_N(t)\}$, and the (unweighted) graph Laplacian matrix $L$, one can obtain a compact form of the complex network system 
\begin{align} \label{eq:complex_network_compact}
    \dot x(t) = f(x(t)) - (K(t)L\otimes I_n) x(t).  
\end{align}

\subsection{A general weight updating law for distributed adaptive synchronization}

In light of Theorem~\ref{theorem:main_gain_updating}, we propose the following distributed adaptive updating law for tuning local  coupling weights
\begin{align} \label{eq:general_updating_law_node_u}
    \dot k_{i}(t) = c_i \left\|\sum_{j \in \mathcal{N}_i} (x_j(t) - x_i(t))\right\|^{p_i}, k_{i}(0)>0,
\end{align}
where $c_i, p_i$ are positive constants with $c_i >0, p_i \geq 1$. 

Note that the graph Laplacian matrix $L$ is a  singular  M-matrix, while the (usually nonlinear and unknown) vector function  $f$ under the QUAD assumption takes a similar role of the unknown matrix $A$ in the uncertain system \eqref{eq:system}. Therefore the complex network model \eqref{eq:complex_network_compact} resembles the System (I) of \eqref{eq:system}, and one can expect that with adaptive and monotonically increasing weights $k_i(t)$ updated by \eqref{eq:general_updating_law_node_u} the network synchronization can be achieved. We formalize this intuition in the following theorem,  with a careful treatment of the singularity of the Laplacian matrix $L$.  
Since the result is of its own interest for network synchronization study,  we   present it as a main theorem with a  proof. 
 
\begin{theorem} \label{theorem:main_result_undirected_node}
Consider the complex network system \eqref{eq:complex_network_compact} with the general distributed  updating laws \eqref{eq:general_updating_law_node_u} that adjust individual coupling weight for each distributed system. Suppose the underlying graph is undirected and connected. Then the following property and convergence results hold true.
\begin{enumerate} 
\item All individual systems of the complex  network \eqref{eq:complex_network_compact} achieve state synchronization globally and asymptotically; furthermore, there exists a finite time $\bar t$ such that the synchronization is achieved exponentially fast $\forall t > \bar t$.  
\item All distributed coupling weights $k_{i}(t)$  are upper bounded for all the time, and converge to some constant values; i.e.,  $\lim_{t \rightarrow \infty} k_{i}(t) \rightarrow k^i_{\infty} <\infty$ for some constant   and bounded value $k^i_{\infty}   >0$, $\forall i$. 
\end{enumerate}
\end{theorem}

\textbf{Proof} 
Construct a Lyapunov-like function 
\begin{align}
    V(x(t)) = \frac{1}{2} (\bar Hx)^T   (\bar Hx) = \frac{1}{2} x^T  \bar L x,
\end{align}
where $\bar H = H \otimes I_n$, $\bar L = L \otimes I_n$, and $H$ is the associated incidence matrix of the graph (with each row corresponding to an edge of the graph under arbitrary orientation assigned). For undirected graphs, the Laplacian matrix can be decomposed as $L = H^TH$ (see e.g., \cite{mesbahi2010graph}). From the QUAD condition \eqref{eq:QUAD} for the vector function $f$   in Definition~\ref{Def:QUAD_condition}, one can obtain
\begin{align}
 (\bar Hx)^T   \bar H f(x) &= \sum_{(i,j) \in \mathcal{E}}(x_i-x_j)^T  \left(f(x_i) - f(x_j) \right)     \nonumber \\
 &\leq \sum_{(i,j) \in \mathcal{E}}(x_i-x_j)^T (\Delta - \epsilon I_n) (x_i-x_j)   \nonumber \\
& = (\bar Hx)^T  \left(I_N \otimes (\Delta - \epsilon I_n) \right) \bar Hx
\end{align}
where $\mathcal{E}$ denotes the edge set of the underlying graph.   The Lie derivative of $V(x(t))$ along the solutions of the complex network system \eqref{eq:complex_network_compact} can be derived as 
\begin{align}
    \dot V(x(t)) =  & (\bar Hx)^T  (\bar H \dot x)    \nonumber \\
    = &(\bar Hx)^T  \left(\bar H (f(x) -   (K(t)L\otimes I_n) x(t) )\right)     \nonumber \\
   = &(\bar Hx)^T   \bar H f(x) - (\bar Hx)^T   \bar H  \bar  K(t) \bar H^T \bar Hx \nonumber \\
   \leq & (\bar Hx)^T (I_N \otimes (\Delta - \epsilon I_n)) \bar Hx \nonumber \\
   &- (\bar Hx)^T   \bar H  \bar  K(t) \bar H^T \bar Hx, 
\end{align}
where $\bar K(t) = K(t) \otimes I_n$. The matrix $H    K(t)   H^T$ is the edge-based weighted Laplacian matrix for the undirected graph. With the monotonic increasing of each weight $k_i(t)$ the non-zero eigenvalues of $H    K(t)   H^T$ also monotonically increase along with time (see Lemma \ref{lemma:edge_laplacian} in Appendix).  Note that it holds $(\bar Hx)^T   \bar H  \bar  K(t) \bar H^T \bar Hx \geq \lambda^+_{\text{min}}(\bar H  \bar  K(t) \bar H^T) \|\bar H x\|^2$ where $\lambda^+_{\text{min}}$ denotes the smallest positive eigenvalue of the associated edge weighted Laplacian (see \cite{sun2018event}). Therefore $\dot V(x(t)) \leq (\bar Hx)^T (I_N \otimes (\Delta - \epsilon I_n) - \lambda^+_{\text{min}}(\bar H  \bar  K(t) \bar H^T)I_{Nn}) \bar Hx$. With the monotonic increasing of the local weight function $k_i(t)$ updated by the state-dependent law \eqref{eq:general_updating_law_node_u}, there must exist a finite time $t^*$, such that $\forall t> t^*$, $\mathcal{A}(t) := (I_N \otimes (\Delta - \epsilon I_n) - \lambda^+_{\text{min}}(\bar H  \bar  K(t) \bar H^T)I_{Nn}) \prec 0$ and therefore $\dot V(x(t)) \leq -\lambda_{\text{min}}(-\mathcal{A}(t))\|\bar Hx\|^2$, $\forall t> t^*$. This again implies that $\|\bar H x\| \rightarrow 0$ with an exponential rate of  at least $\lambda_{\text{min}}(-\mathcal{A}(t^*)), \forall t>t^*$. Since the underlying graph is connected which implies  $\text{null}(\bar H) = \text{span} ({\bf{1}}_N \otimes I_n)$ \cite{mesbahi2010graph}, the convergence $\|\bar H x\| \rightarrow 0$ is equivalent to that   $x_i(t) \rightarrow x_j(t), \forall (i,j)$ exponentially fast $\forall t>t^*$, i.e., the state synchronization is achieved asymptotically, and after a finite time $t^*$ the synchronization convergence is exponentially fast. 

Following a similar argument as in the proof of Theorem~\ref{theorem:main_gain_updating}, the exponential convergence of   $\|\bar H x\|$ after a finite time $t^*$ also implies that each $\|\sum_{j \in \mathcal{N}_i} (x_j(t) - x_i(t))\|$ exponentially converges to zero $\forall t> t^*$, and an integral of the  adaptive coupling weight law \eqref{eq:general_updating_law_node_u} ensures the existence of an upper bound of $k_i(t)$ $\forall i, t>0$. Since each $k_i(t)$ is continuous and monotonically increasing,   one concludes that each $k_i(t)$ must be convergent in the limit.  
\qed

\subsection{Discussions:  new insights on distributed adaptive synchronization based on distributed adaptive stabilization theory} 
The distributed adaptive stabilization theory developed in previous sections provides a unified and general framework to study adaptive synchronization. In the following remarks, we present some novel insights on distributed design of coupling weights in complex network synchronization.  

\begin{remark}
The updating law \eqref{eq:general_updating_law_node_u} also includes the following quadratic form as a special case (i.e., setting $p_i = 2$ in \eqref{eq:general_updating_law_node_u})
\begin{align} \label{eq:general_updating_law_node_quadratic}
   & \dot k_{i}(t) =  c_i \left(\sum_{j \in \mathcal{N}_i} (x_j(t) - x_i(t))\right)^T \sum_{j \in \mathcal{N}_i} (x_j(t) - x_i(t)), \nonumber \\
  &  k_{i}(0)>0.
\end{align}

The above quadratic function on the right-hand side of \eqref{eq:general_updating_law_node_quadratic} and its variations are the most popular weight updating law for adaptive synchronization control,  which has been extensively studied  in the literature on adaptive synchronization or consensus control of complex networks (see e.g., \cite{yu2012distributed,li2013consensus, li2013distributed,shafi2015adaptive}). The proof for the adaptive synchronization in these papers often involves a Lyapunov function, in the form of $V = \frac{1}{2} x^T  \bar L x + \sum_{i=1}^N    (\alpha_{i}-k_{i})^2$ (or  similar forms) with  some sufficiently large but unknown $\alpha_{i}$. The stability analysis employs a Lyapunov-based argument and Barbalat's lemma to prove the convergence of the synchronized states. As   demonstrated above one can prove the stability and synchronization convergence of the complex network \eqref{eq:complex_network_compact} with a unified approach for a general weight-updating law \eqref{eq:general_updating_law_node_u}. In addition, via the insights of distributed matrix high gains  and the M-matrix property one can also extend the adaptive synchronization to directed networks, and characterize the exponential convergence of the state synchronization, which is not available by using the conventional approach with Barbalat's lemma as in \cite{li2013distributed,shafi2015adaptive}. Furthermore, the updating law of coupling weights in the form of \eqref{eq:general_updating_law_node_u} generalizes the main result in \cite{delellis2009novel} under much weaker conditions (while it is assumed in \cite{delellis2009novel} that  the QUAD condition should satisfy $\Delta - \epsilon I_n <0$ to ensure adaptive synchronization). 
\end{remark}
 
\begin{figure}[t]
\begin{center}
\includegraphics[width=0.50\textwidth]{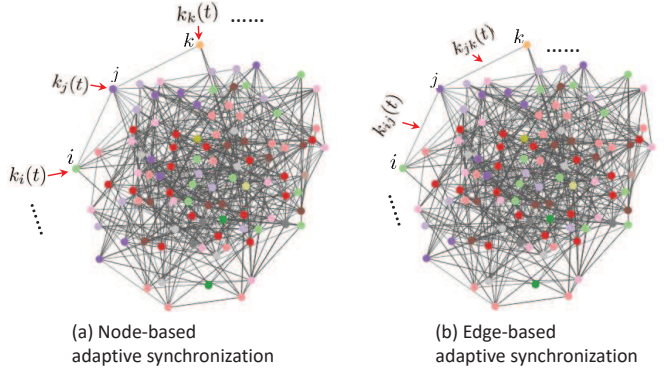}
\caption{Illustrations of adaptive synchronization in complex networks. (a) Node-based adaptive synchronization of the system dynamics \eqref{eq:node_network_dynamics}, corresponding to  the uncertain  system (I) in \eqref{eq:system_example_I}. (b) Edge-based adaptive synchronization of the system dynamics \eqref{eq:edge_network_dynamics}, corresponding to  the uncertain  system (II) in \eqref{eq:system_example_II}.  }
\label{fig:Network}
\end{center}
\end{figure}

\begin{remark}
The network synchronization dynamics in  \eqref{eq:node_network_dynamics}  are often termed \textbf{node-based} adaptive synchronization in the literature, as the updating of adaptive coupling weights in  \eqref{eq:general_updating_law_node_u} is implemented by each individual node system. We remark that the node-based network system \eqref{eq:complex_network_compact} resembles the uncertain linear system (I) in \eqref{eq:system_example_I}, where Theorem~\ref{theorem:main_gain_updating} applies. In contrast, one can also consider the following \textbf{edge-based} adaptive synchronization dynamics (e.g., \cite{yu2012distributed,li2013consensus})
\begin{align} \label{eq:edge_network_dynamics}
    \dot x_i(t) = f(x_i(t)) + \sum_{j \in \mathcal{N}_i}  k_{ij}(t) (x_j(t) - x_i(t)),
\end{align}
where $k_{ij}(t)$ is a time-varying coupling weight function for edge $(i,j)$. An illustration of the node-based and edge-based adaptive synchronization in complex networks is shown in Fig.~\ref{fig:Network}. By numbering the weight function for each edge as $w_l(t) := k_{ij}(t)$ with the same ordering of the graph topology and defining $W(t) = \text{diag}\{w_1(t), \cdots, w_{M}(t)\}$,  the weighted graph Laplacian matrix is described by $L = H^T W H$. In this way
one can obtain a compact form of the complex network system \eqref{eq:edge_network_dynamics} by $ 
    \dot x(t) = f(x(t)) - (H^T W \otimes I_n) (\bar H x(t))
$, which resembles the structure of System (II) in \eqref{eq:system_example_II}. In  light of Theorem~\ref{theorem:main_gain_updating_BK}, we  propose the following distributed \textit{edge-based} updating law for adaptive edge coupling weights
\begin{align} \label{eq:general_updating_law_edge_u}
    \dot k_{ij}(t) = c_{ij} \left\| (x_j(t) - x_i(t))\right\|^{p_{ij}}, k_{ij}(0)>0,
\end{align}
where $c_{ij} >0, p_{ij} \geq 1$. Similarly, by following Theorem~\ref{theorem:main_gain_updating_BK}, one can expect that under the adaptive weights \eqref{eq:general_updating_law_edge_u} the network system \eqref{eq:edge_network_dynamics} will achieve state synchronization while all edge weights $k_{ij}(t)$ converge to bounded values. A detailed study of general edge-based weight coupling laws for adaptive synchronization can be found in \cite{wang2019adaptive}. 
\end{remark}

\subsection{Numerical examples}
In the simulations we consider a complex network consisting of 100 Van  der  Pol oscillators coupled by an undirected graph. Each node in the network represents a  Van  der  Pol oscillator, whose system dynamics can be described by the following second-order nonlinear equation (see e.g., \cite{wang2005partial})
 \begin{align}
     \dot x_i &= wy_i - \frac{a}{3}x_i^3 - bx_i + k_i(t)\sum_{j\in \mathcal N_i} (x_j - x_i),  \nonumber \\
     \dot y_i &= -wx_i+\frac{\mu(t)}{w}  +  k_i(t)  \sum_{j\in \mathcal N_i} (y_j - y_i),
 \end{align}
 where $x_i, y_i$ are the states of the $i$-th oscillator, $w, a, b$ are system parameters, $\mu(t)$ is a  driven force term, and $k_i(t)$ is a local adaptive coupling weight for the $i$-th oscillator. In the simulations, we set the parameters as $a = w = b = 1$ and $\mu(t) = \text{sin}(t)$ which are the same as in \cite{delellis2011quad}. Some upper bounds of the QUAD conditions for the  Van  der  Pol oscillator  with the same parameters are estimated in \cite{delellis2011quad}. We remark that  adaptive synchronization does not require to know any true or estimated values of these bounds. 

 \begin{figure*}[t]
\begin{center}
\includegraphics[width=0.7\textwidth]{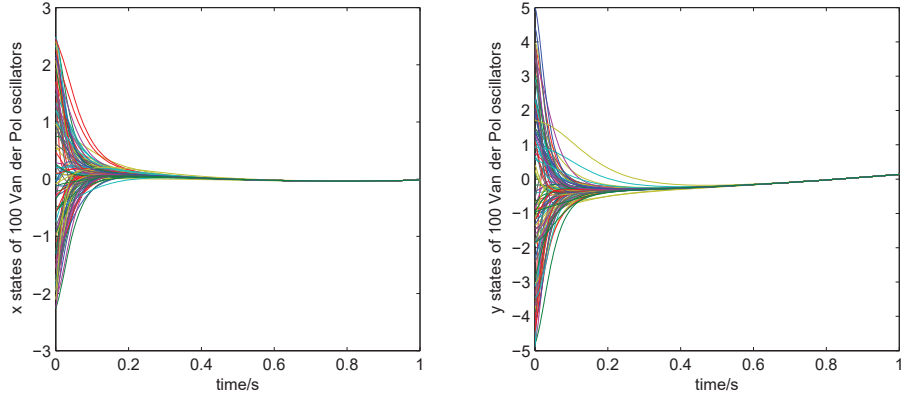}
\caption{State synchronization of 100 Van der Pol oscillators connected by an Erd{\H{o}}s--R{\'e}nyi   network with adaptive distributed  coupling weights. Left: synchronization of the $x$ states. Right: synchronization of the $y$ states.  }
\label{fig:state_synchronization}
\end{center}
\end{figure*}
 
\begin{figure}[t]
\begin{center}
\includegraphics[width=0.5\textwidth]{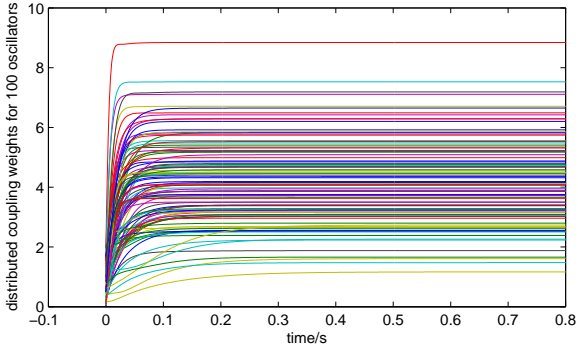}
\caption{Time  evolution of local adaptive coupling weights for 100 distributed oscillators.}
\label{fig:coupling_weights}
\end{center}
\end{figure}

We model the oscillator coupling graph by the Erd{\H{o}}s--R{\'e}nyi network  \cite{van2016random}, and generate an Erd{\H{o}}s--R{\'e}nyi random graph $G(n, \rho)$ with $n = 100$ nodes and by a connectivity probability $\rho = 0.10$ (an illustration is shown in Fig.~\ref{fig:Network}(a)).   Each Van  der  Pol  oscillator in each node adaptively updates its own local coupling weight via \eqref{eq:general_updating_law_node_u} with relative information received from its neighbors in order to achieve state synchronization. In the simulations, the initial states of all oscillators are randomly chosen, while the initial coupling weights are also selected to be random positive values. The  parameters in the weight updating law are set  as $c_i = 1$ with a power parameter $p_i = 1.5, \forall i$ according to \eqref{eq:general_updating_law_node_u}.  It can be seen from Fig.~\ref{fig:state_synchronization} that the states of all oscillators are asymptotically synchronized. All distributed coupling weights are updated by local information from   neighboring oscillators  to ensure synchronization, which all converge to their individual bounded values as shown in Fig.~\ref{fig:coupling_weights}.

\section{Conclusions} \label{sec:conclusions} 
In this paper we have presented a distributed adaptive stabilization theory for uncertain multivariable systems with matrix high gains, while the adaptive gains are described by a time-varying positive diagonal matrix. Adaptive matrix gain stabilization is motivated by distributed and scalable stabilization of spatially distributed systems, and will find many applications in  distributed control for networked and coupled systems. We show that the unknown system matrix $B$ being an H-matrix with positive diagonal entries guarantees matrix high-gain stabilizability of uncertain multivariable systems. We propose a general approach for designing state-dependent updating laws for individual gain functions, and prove the  convergence of both  system states and  adaptive matrix gains. 
We show an application of adaptive synchronization for complex network systems, while each node dynamics can update its own and local coupling weights to ensure state synchronization. Based on the distributed adaptive stabilization approach, a unified framework of network synchronization control is proposed which suggests several general and novel designs for node-based and edge-based local updating laws of coupling weights to achieve adaptive network synchronization. 

\section{Appendix}
\subsection{Matrix measure}
The matrix measure (or ``logarithmic norm") plays an important role in bounding the solution of differential equations.  We introduce the definition and some properties of matrix measure from \citep{desoer1975feedback} as follows.

\begin{definition}(Matrix measure)
Given a real $n \times n$ matrix  $A$, the  matrix measure   $\mu(A)$ is defined as 
\begin{align}
    \mu(A) = \text{lim}_{\epsilon \downarrow 0} \frac{\|I + \epsilon A\| - 1}{\epsilon},
\end{align}
where $\|\cdot\|$ is a matrix norm on $\mathbb{R}^{n \times n}$ induced by a vector norm $\|\cdot\|'$ on $\mathbb{R}^n$. 
\end{definition}
The matrix measure is always well-defined, and can take positive or negative values. Different matrix norms  on $\mathbb{R}^{n \times n}$ induced by a corresponding  vector norm $\|\cdot\|'$ give  rise to different matrix measures. In particular, one can show the following two commonly-used matrix measures. 
\begin{itemize}
    \item If the vector norm $\|\cdot\|'$ is chosen as the 1-norm, i.e.,  $\|\cdot\|' = \|\cdot\|_1$, then the induced matrix norm is the column-sum norm, i.e., $\|A\| = \|A\|_{\text{col}} = \text{max}_j \sum_i |a_{ij}|$. The corresponding matrix measure is 
    \begin{align} \label{eq:measure_one_norm}
        \mu(A) = \text{max}_{j =1, 2, \cdots, n} \left(a_{jj} +\sum_{i=1, i\neq j}^n |a_{ij}|\right).
    \end{align}
    
        \item If the vector norm $\|\cdot\|'$ is chosen as the $\infty$-norm, i.e., $\|\cdot\|' = \|\cdot\|_\infty$, then the induced matrix norm is the row-sum norm, i.e., $\|A\| = \|A\|_{\text{row}} = \text{max}_i \sum_j |a_{ij}|$. The corresponding matrix measure is 
    \begin{align} \label{eq:measure_infinity_norm}
        \mu(A) = \text{max}_{i =1, 2, \cdots, n} \left(a_{ii} +\sum_{j=1, j\neq i}^n |a_{ij}|\right).
    \end{align}
\end{itemize}

\subsection{Solution bounds of time-varying linear systems}
We recall the following result (the Coppel inequality \cite{coppel1965stability}) that bounds the solution of a time-varying linear system via matrix measures (see e.g., Chapter 2 of \citep{desoer1975feedback}). 

\begin{lemma} \label{theorem:measure_exponential}
Let $t \rightarrow A(t)$ be a continuous matrix function from $\mathbb{R}^+$ to $\mathbb{R}^{n \times n}$. Then the solution of the time-varying linear system 
\begin{align}
    \dot x(t) = A(t)x(t)
\end{align}
satisfies the inequalities
\begin{align}
    \|x(t_0)\|' e^{- \int_{t_0}^{t} \mu(-A(t'))\text{d}t'}  \leq \|x(t)\|' &  \leq \|x(t_0)\|' e^{\int_{t_0}^{t} \mu(A(t'))\text{d}t'}, \nonumber \\
    & \forall t \geq t_0
\end{align}
where $\|\cdot\|'$ denotes a vector norm that is compatible with the norm in the matrix measure $\mu(A)$. 
\end{lemma}

\subsection{Proof of Theorem~\ref{theorem:_H_matrix}}
\textbf{Proof}
The equivalence of each statement is proved as below. 
\begin{itemize}
    \item $(i) \iff (ii)$ \\
    This is a reformulation of the statement  in Lemma~\ref{lemma:H_matrix}.
    \item $(ii) \iff (iii)$ \\
    Under a positive diagonal matrix $\bar D$, the entries of the matrix $\bar A = \{\bar a_{ij}\} := \bar D^{-1} A \bar D$ are described by 
    \begin{align}
    \bar a_{ij} =   \left\{
       \begin{array}{cc}
       a_{ij},  &\text{  if  } \,\,\,\,j  = i;  \\ \nonumber
       \frac{\bar d_j}{\bar d_i}a_{ij},  &\text{  if  } \,\,\,\, j  \neq i.   \nonumber  
       \end{array}
      \right.
\end{align}
    By definition, strict  row-diagonal dominance of $\bar D^{-1} A \bar D$ equivalently indicates that 
    \begin{align}
    |\bar a_{ii}| = |a_{ii}| &> \sum_{j=1, j \neq i}^{n}\bar a_{ij} \nonumber \\
    &= \frac{1}{\bar d_i} \sum_{j=1, j \neq i}^{n} \bar d_j |a_{ij}|, \forall i = 1, 2, \cdots, n,
\end{align}
    which is equivalent to  $
    |a_{ii}| \bar d_i> \sum_{j=1, j \neq i}^{n} \bar d_j |a_{ij}|, \forall i = 1, 2, \cdots, n$ since $\bar d_i >0, \forall i$. The latter inequality equivalently implies that $A$ is generalized row-diagonally dominant according to Definition~\ref{def:gene_row_DD}. 
    Therefore the equivalence is proved.   
    \item $(i) \iff (iv)$ \\ 
    We first show that $A$ is an H-matrix if and only if $A^T$ is an H-matrix. 
    According to Definition~\ref{def:M_matrix} and properties of M-matrix \cite{roger1994topics}, a matrix $M_A$ is an M-matrix if and only if that there exists a positive scalar  $s>0$ and a non-negative matrix $N = \{n_{ij} \geq 0\} \in \mathbb{R}^{n \times n}$, such that $M_A = sI_n - N$ and $s> \rho(N)$. Without loss of generality we choose $s = \text{max} \{a_{ii}, i = 1,2,\cdots, n\}$, and therefore $N = sI_n -M_A$ which is a non-negative matrix. Since $M_A^T = (sI_n -N)^T = sI_n -N^T$ and because $\rho(N) = \rho(N^T)$, one has $s > \rho(N^T)$ and therefore $M_A^T$ is also an M-matrix if $M_A$ is an M-matrix. By Definition~\ref{def:comparison_matrix} and the structure of the comparison matrix, one concludes that $A$ is an H-matrix if and only if $A^T$ is an H-matrix. Then applying statement (ii) to $A^T$ gives the result. 
    \item $(iv) \iff (v)$ \\ 
    Under  a positive diagonal matrix $\tilde D$, the entries of the matrix $\tilde A = \{\tilde a_{ij}\} = \tilde D A \tilde D^{-1}$ are described by 
    \begin{align}
    \tilde a_{ij} =   \left\{
       \begin{array}{cc}
       a_{ij},  &\text{  if  } \,\,\,\,j  = i;  \\ \nonumber
       \frac{\tilde d_i}{\tilde d_j}a_{ij},  &\text{  if  } \,\,\,\, j  \neq i.   \nonumber  
       \end{array}
      \right.
\end{align}
    By definition, strict column-diagonal dominance of $\tilde D A \tilde D^{-1}$ equivalently indicates that 
    \begin{align}
    |\tilde a_{jj}| = |a_{jj}| &> \sum_{i=1, i \neq j}^{n} \tilde a_{ij}  
    \nonumber \\
    &= \frac{1}{\tilde d_j} \sum_{i=1, i \neq j}^{n} \tilde d_i |a_{ij}|, \forall j = 1, 2, \cdots, n,
\end{align}
which is equivalent to $|a_{jj}| \tilde d_j>   \sum_{i=1, i \neq j}^{n} \tilde d_i |a_{ij}|, \forall j = 1, 2, \cdots, n$ since $\tilde d_j >0, \forall j$.  The latter inequality equivalently implies that $A$ is generalized column-diagonally dominant according to Definition~\ref{def:gene_column_DD}. 
Therefore the equivalence is proved. 
\end{itemize}  \qed

\subsection{A lemma on weighted edge Laplacian}
The following lemma shows the monotonic increasing property of non-zero eigenvalues of weighted edge Laplacian with monotonic increasing node weights.

\begin{lemma} \label{lemma:edge_laplacian}
Consider two undirected connected graphs $\mathcal{G}_{K}$ and $\mathcal{G}_{\hat K}$ with the same node-edge topology (encoded by the 0-1 incidence matrix $H \in \mathbb{R}^{m \times n}$), but with different sets of positive node weights $K = \text{diag}\{k_1, k_2, \cdots, k_n\}$ and $\hat K = \text{diag}\{\hat k_1, \hat k_2, \cdots, \hat k_n\}$. Their weighted edge Laplacian matrices are denoted by $L_{\mathcal{G}_{K}}: = H K H^T$ and $L_{\mathcal{G}_{\hat K}}: = H \hat K H^T$, respectively, and their eigenvalues are listed in an ascending order $\lambda_{i}(\mathcal{G}_{K})$ and $\lambda_{i}(\mathcal{G}_{\hat K})$, $\forall i =1,2, \cdots, m$, respectively.  
\begin{itemize}
    \item The two matrices have the same number of zero eigenvalues: $\lambda_{i}(L_{\mathcal{G}_{K}}) = \lambda_{i}(L_{\mathcal{G}_{\hat K}}) =0$, $i = 1, \cdots, \kappa$, where $\kappa$ is the number of independent cycles in the graph.  
    \item  If the node weights of the two weighted graphs satisfy  $k_{i}  > \hat k_{i}$, $\forall i =1,2, \cdots, n$, then the non-zero eigenvalues of the weighted edge Laplacians satisfy 
\begin{align}
\lambda_{i}(L_{\mathcal{G}_{K}}) > \lambda_{i}(L_{\mathcal{G}_{\hat K}}),\,\,\,\forall  i  = \kappa +1, \cdots, m.
\end{align}
In particular, $\lambda_{min}^+ (  H   K   H^T) > \lambda_{min}^+ (  H   \hat K   H^T) >0$. 
\end{itemize}

\end{lemma}

\textbf{Proof}  
For a connected undirected graph $\mathcal{G}$ encoded by the incidence matrix $H$ (with arbitrary
orientation assigned), the null space $\text{null}(H^T)$ is spanned by all  linearly independent signed path vectors corresponding to
the cycles in $\mathcal{G}$ \cite{godsil2013algebraic}. Then the dimension of $\text{null}(H^T)$ is the number of independent cycles in the graph. Since the diagonal weight matrices $K$ and $\hat K$ are positive definite,   one has $\text{null}(H^T) = \text{null}(H K H^T) = \text{null}(H \hat K H^T)$, from which one concludes that the two edge Laplacian matrices $L_{\mathcal{G}_{K}}$ and $L_{\mathcal{G}_{\hat K}}$ have the same number $\kappa$ of zero eigenvalues. 

  Now we analyze the non-zero eigenvalues of the edge Laplacian matrices.
By denoting $\tilde {k}_i = k_{i} - \hat k_{i} >0, \forall i$ and $\tilde K = \text{diag}\{\tilde {k}_1, \tilde {k}_2, \cdots, \tilde {k}_n\}$, one has $L_{\mathcal{G}_{K}} = H (\hat K + \tilde K) H^T = L_{\mathcal{G}_{\hat K}} + H\tilde K H^T$. The matrix $H\tilde K H^T$ has the same number of zero eigenvalues  with the same null vectors   as in $L_{\mathcal{G}_{K}}$, and all other eigenvalues are positive. Therefore, one concludes $\lambda_{i}(L_{\mathcal{G}_{K}}) > \lambda_{i}(L_{\mathcal{G}_{\hat K}}),\,\,\,\forall i = \kappa +1, \cdots, m$. In particular, for the smallest positive eigenvalue, there holds $\lambda_{min}^+ (  H   K   H^T) > \lambda_{min}^+ (  H   \hat K   H^T) >0$. 
\qed

\begin{ack}                               
The work was supported by the Swedish Research Council, the European Research Council (AdG 834142), the Wallenberg AI, Autonomous Systems and Software Program, the Swedish Research Council and the Swedish Foundation for Strategic Research, Sweden (RIT15-0038), National Natural Science Foundation of China under Grant 61973006, and Beijing Natural Science Foundation under grant JQ20025. The authors would like to thank Prof. Karl Johan {\AA}str{\"o}m for helpful discussions on adaptive stabilization. The work of Zhiyong Sun is supported by a starting grant from Eindhoven Artificial Intelligence Systems Institute (EAISI).  
\end{ack}


\bibliographystyle{IEEEtran}
\bibliography{adaptive_synchronization}         

\end{document}